# MOVING BEYOND PRINCIPLES: IDENTIFYING ACTIONABLE AI FAIRNESS PRACTICES

*Completed Research Paper*

Christoph Burtscher, University of Reading, Reading, United Kingdom, c.burtscher@reading.ac.uk

Mateusz Dolata, Zeppelin University, Friedrichshafen, Germany, mateusz.dolata@zu.de

## Abstract

*Because artificial intelligence (AI) increasingly mediates organizational work, fairness has become a critical governance challenge. Existing frameworks often prioritize abstract ethical principles rather than fairness-specific ones and lack actionable guidance across the entire AI lifecycle. This study addresses the principles-to-practice gap in AI fairness governance. We develop actionable AI fairness practices and draw on a socio-technical and praxiological lens, conducting discourse and thematic analyses of 60 academic, policy, and practitioner sources. From these analyses, we derive a structured set of AI fairness practices in a comprehensive, AI lifecycle-spanning matrix organized by obligation degree and organizational role. The matrix provides dynamic, role-specific guidance to support implementation and sustainment of AI fairness. By extending the AI fairness beyond abstract principles to operationalized, actionable practices, we contribute to IS scholarship and offer a modular governance scaffold.*

## 1 Introduction

Artificial intelligence systems increasingly shape critical organizational decisions that affect employees, customers, and society broadly. However, their deployment often perpetuates systemic discrimination rather than eliminating it (Dolata et al., 2022). Organizations that implement AI face growing pressure to ensure fairness. This is not just an aspiration but an operational imperative connected to regulatory compliance, stakeholder trust, and sustainable value creation (Mäntymäki et al., 2023). IS has framed AI fairness as a sociotechnical challenge that extends beyond algorithmic approaches to bias mitigation. It also encompasses governance structures, organizational practices, and the interplay between human actors and technological artifacts (Dolata et al., 2022; Dolata & Schwabe, 2023).

AI fairness is widely considered a dimension of responsible AI. Common to various definitions of responsible AI are accountability, transparency, explainability, privacy, security, and human control (De Almeida et al., 2021). Despite intensive discourse around responsible AI and its implementation, organizations struggle to translate high-level ethical principles into actionable governance practices (Ali et al., 2023; Horneber, 2025; Mittelstadt, 2019). This principles-to-practice gap manifests in three critical ways. First, existing AI governance frameworks remain insufficiently operationalized for practitioners, offering abstract principles that organizational actors cannot readily implement in their daily work (Morley et al., 2023; Schiff et al., 2020). Companies face challenges translating ethical principles into practice, with methods and tools often unused because they fail to meet practitioner needs or remain unknown to those who require them (Mäntymäki et al., 2023). Second, guidance on AI fairness is fragmented across disciplines such as computer science, philosophy, law, and management, leading to conflicting recommendations that organizations fail to synthesize into coherent governance systems (Dolata et al., 2022). Third, current approaches emphasize risk mitigation rather than value creation, focusing on protecting against discrimination and meeting regulatory criteria while overlooking opportunities to maximize fairness as a strategic capability of an organization (Davenport & Redman, 2025).

These limitations are particularly acute for AI fairness, which differs fundamentally from some other responsible AI dimensions. Unlike privacy or security, which can be primarily addressed through technical controls and compliance mechanisms, fairness involves inherently social constructs that sometimes have an emergent character rather than being clearly specified or operationalized (Dolata & Schwabe, 2023; Ochmann et al., 2024). Fairness perceptions are not universal but socially constructed (Dolata & Schwabe, 2023), context-dependent, and dynamic over time, requiring governance approaches that accommodate this complexity (Bargh et al., 2025). We claim that without understanding the public discourse and factors shaping fairness expectations, scholars cannot offer practical guidance to organizations and policymakers on implementing and maintaining AI fairness.

This study addresses these research gaps through a phenomenon-based problematization approach grounded in praxiological and sociotechnical lenses (Dolata et al., 2022). Rather than proposing yet another set of abstract principles, we leverage discourse analysis of academic literature, policy documents, and industry reports to identify the practices—not merely principles—that organizations can implement to govern for AI fairness (Schiff et al., 2020). By adopting a praxiological lens that emphasizes human activity and recognizing AI systems as complex sociotechnical configurations involving actors, technologies, tasks, and structures, we offer actionable practices spanning the entire AI lifecycle.

Therefore, this research asks: *How can organizations define, design, and implement AI governance to ensure AI fairness?*.

By synthesizing fragmented guidance into a unified, fairness-specific set of practices that are dynamic, contextually tailored, and pragmatic in their obligation structure, this study contributes to IS scholarship on AI governance. It provides organizations with practices for achieving algorithmic justice.

The paper proceeds as follows: Section 2 provides the theoretical foundations and introduces AI, AI fairness, and governance. Section 3 outlines the methodology for the discourse and thematic analysis. Section 4 presents the empirical findings with a thematic map and AI fairness practices. Section 5 provides a summary of the main conclusions, discusses the implications, and the limitations of this study.

## 2 Theoretical Foundations

### 2.1 AI Fairness

Fairness has emerged as a critical issue due to its potential impact on various stakeholders, including employees, customers, and society. Fairness includes "treating others as they should or deserve to be treated by adhering to standards of right and wrong" (Teodorescu et al., 2021, p. 1019). Ethical AI ensures human well-being (Dignum, 2018) and incorporates AI fairness as a key principle (Jobin et al., 2019). Discriminatory non-harm based on equality and non-discrimination (Ryan & Stahl, 2021) for equal treatment and outcomes is central to AI fairness (Leslie et al., 2023). Biases underlie unfairness, and fairness mitigates their effects (Li et al., 2023).

Fairness and its perceptions are often studied through the lens of organizational justice (Li et al., 2014). It refers to "the perception and understanding of fairness in organizations" (Kals & Jiranek, 2012, p. 222). It provides the context for AI fairness practices in organizations, which are primarily focused on distributive non-discrimination and procedural equality rather than justice, with the following four goals. According to the organizational justice theory (Greenberg, 1986), justice includes procedural, distributive, interpersonal, and informational fairness (Colquitt, 2012; Greenberg, 1990; Robert et al., 2020). Distributive fairness focuses on content and emphasizes "fairness of the ends achieved" (Greenberg, 1990, p. 400). Procedural focuses on process and is focused on the "fairness of means used to achieve those ends" (Greenberg, 1990, p. 400). Interactional justice is differentiated into interpersonal and informational justice (Colquitt et al., 2005). Interpersonal justice refers to "showing concern for individuals regarding the distributive outcome they receive" (Greenberg, 1993, p. 85), representing respect and propriety (Colquitt et al., 2005). Informational justice refers to "providing knowledge about procedures that demonstrate regard for people's concerns" (Greenberg, 1993, p. 84), representing truthfulness and justification (Colquitt et al., 2005). The activities of organizations in achieving AI fairness aim to fulfill justice goals (Binns, 2018). Whereas this already indicates the complex nature of fairness and perceptions thereof, the picture becomes even more complex when we consider that notions of fairness vary across situations and may undergo significant shifts over time (Bargh et al., 2025; Dolata & Schwabe, 2023; Ochmann et al., 2024).

Algorithms, which were initially praised as potential ways to achieve impartial, bias-free treatment of citizens or clients, turned out to sustain, amplify, and reinforce past prejudices (Christian, 2021). This necessitated that for AI and machine learning communities to devise ways to mitigate systematic biases by introducing additional filters and mathematical evaluations (Feuerriegel et al., 2020). Hence, algorithmic fairness can be seen as a technology that prevents systematic biases in algorithms. The debate over whether those measures are sufficient continues. Part of the problem is the numerous, mutually exclusive notions of mathematical fairness (Mitchell et al., 2021). Determining which one is applicable in which context makes the problem and potential answers more precise but not necessarily more accurate. Choosing which measure is appropriate and when remains a social and moral dilemma that technology cannot easily solve (Christian, 2021; Mitchell et al., 2021).

However, the issue of technological justice should not be limited to inadequate or biased decisions. Apart from the distribution of goods or rights, justice also pertains to whether the procedure in which decisions were made was appropriate and whether a person was treated adequately, given their behavior (Dolata et al., 2022; Greenberg, 1986; Robert et al., 2020). Justice restoration, i.e., treating someone who was treated unfairly in the past "whole" again, can be achieved by granting them more rights than other members of the population. Similarly, retributive justice, i.e., punishing someone who has caused the injustice, can be achieved by depriving other members of the population (Bradfield & Aquino, 1999). Accordingly, a treatment that appears unjust in a narrow sense might be adequate in a specific historical context. This suggests that justice and perceptions of justice are not limited to the context of a single decision but rather relate to various treatment attributes and the subject of the decision. This is particularly important for organizations that need to ensure consistent, fair treatment of employees.

Most studies concerning AI fairness refer implicitly or explicitly to the distributive justice (Dolata et al., 2022; Starke et al., 2022). However, some studies go beyond this scope. Notable examples include Decker et al. (2025) and Marcinkowski et al. (2020). While the number of empirical studies in this context is growing, the available guidance on implementing AI fairness frequently abstracts from the specific dimension of justice. At the same time, evidence suggests that different aspects (e.g., explanations, consistency) might interact differently with varying dimensions of justice (Starke et al., 2022).

Overall, fairness is one of the core dimensions of responsible and trustworthy AI, alongside human oversight, transparency, privacy, auditability, and traceability (Lu et al., 2024). At the same time, we argue that treating those aspects alike is dangerous. As for fairness, aspects such as its subject to social negotiation (Dolata & Schwabe, 2023) or inherent ambiguity (Chowdhury & Klautzer, 2025) differentiate it from other responsible AI domains. Additionally, the complex nature of fairness perceptions, as explicated by its dimensions, warrants attention. Overall, it calls for a more adaptive, dynamic, and differentiated approach towards AI fairness and its governance (Chowdhury & Klautzer, 2025).

## 2.2 Governance

Corporate governance provides the overarching framework for IT governance, which governs technology and AI (Mäntymäki et al., 2022). The system by which organizations are directed and controlled represents corporate governance (Cadbury, 1992). Organizations manage AI and utilize governance. Governance is defined as "specifying the framework for decision rights and accountabilities to encourage desirable behavior in the use of IT" (Weill, 2008, p. 3). Desirable IT behavior contributes to achieving organizational goals. Governance enables organizations to ensure alignment between IT-related activities and the organization's strategy and objectives (Wu et al., 2015). Effective use of technology relies on good IT governance (Wu et al., 2015).

The normative, action-oriented definition of Mäntymäki et al. (2022, p. 604) defines AI governance as a "system of rules, practices, processes, and technological tools that are employed to ensure an organization's use of AI technologies aligns with the organization's strategies, objectives, and values; fulfills legal requirements; and meets principles of ethical AI followed by the organization.".

Various frameworks are available for corporate or IT governance, but few focus on AI Governance (Wirtz et al., 2020). AI governance is part of IT governance, which in turn is part of corporate governance, and overlaps with data governance in terms of data management (Mäntymäki et al., 2022). As AI increases in complexity and risks, AI systems become opaque and unpredictable, and existing governance frameworks are ill-equipped; governance of AI is required (Taeihagh, 2021). AI governance must address the entire lifecycle of AI, from design and development to use and operation (Eitel-Porter, 2021), and assist in managing issues arising from AI (Butcher & Beridze, 2019).

Guidelines and principles are often utilized to express governance (Jobin et al., 2019). Guidelines refer to guidance on how to address a particular issue (Ryan & Stahl, 2021) and are persuasive (Jobin et al., 2019). Principles, on the other hand, act as guides to action (Whittlestone et al., 2019) and aspirations.

## 2.3  AI Fairness Governance

This paper focuses on AI fairness and its governance. Fairness is a key component of responsible AI. Based on Mäntymäki et al. (2022) AI governance definition, we define **AI fairness governance** as a system of rules, practices, processes, and technological tools that are employed to ensure an organization's use of AI technologies aligns with the organization's strategies, objectives, and values; fulfills legal requirements; and meets principles of fair AI followed by the organization. We refer to AI within the context of AI fairness governance as the object of governance, i.e., the governance of AI, not as the subject or as AI used to govern. The focus is on AI governance for fairness, not on data, IT, or corporate governance. AI Fairness in the context of socio-technical systems (STS) involves not only technology, including AI systems that focus on algorithmic fairness, but also, more comprehensively, the social system components of human actors and structures, as well as the technical system components of technology, including AI, and tasks (Bostrom & Heinen, 1977).

Governance frameworks in AI encompass the AI regulatory and governance framework (De Almeida et al., 2021), which focuses on the interplay between legislative, executive, and judicial government institutions, as well as society, academia, industries, and relevant committees. Their interactions incorporate various AI governance models (Papagiannidis et al., 2022), focusing on the structural, procedural, and relational influences of AI governance, with inhibitors and enablers shaping specific outcomes.

However, there is a nascent and emerging practice trend in responsible AI that shifts attention from high-level principles and metrics to situated practices in organizational contexts and socio-technical arrangements, as exemplified by (Baldassarre et al., 2024; Bateni et al., 2022; Lu et al., 2024; Lu et al., 2023; Mäntymäki et al., 2022) recent works. Such studies intend to translate norms into operations, pattern-oriented engineering approaches, sociotechnical maturity criteria, and operationalizations. The underlying claim is that responsible AI outcomes depend on what actors actually do with tools, rules, and structures over time rather than on declarations of values or isolated model metrics (Barletta et al., 2023). Although this approach is essential for moving from discourse to organizational implementation, the shortage of dedicated studies on practices explicitly related to AI fairness and its governance is problematic. Whereas generic attention to responsible AI as a whole is a good starting point, we argue that the highly social and dynamic nature of fairness merits a more dedicated approach.

Current AI fairness governance frameworks exhibit five critical limitations. First, their *scope* remains overly generic (cf. Baldassarre et al. (2024)), often anchored in broad AI ethics rather than fairness-specific concerns (Barletta et al., 2023), which constrains practical applicability. Additionally, some guidance focuses on the fairness of the overall system components (such as Bateni et al. (2022) for data fairness). Second, *guidance* is dominated by principle-based prescriptions and the proliferation of principles (Floridi & Cowls, 2021) that lack actionable detail tailored to specific stakeholders, leaving practitioners without concrete implementation pathways. Third, *coverage* is fragmented, focusing on discrete software development phases rather than addressing fairness across the entire AI lifecycle in integrated ways, leading to siloed practices and inefficiencies. Fourth, these frameworks lack *dynamism*, assuming static notions of fairness despite evidence that fairness perceptions are socially constructed and can be shaped by algorithmic decision-making in organizational contexts. Finally, the *application* is weak; while structural models exist, they fail to provide context-sensitive, operational strategies, thereby limiting the utility of governance frameworks. Collectively, these deficiencies point to a principles-to-practice gap in the scientific discourse and impede organizations' ability to translate normative aspirations into practice-oriented activities to achieve AI fairness.

# 3 Discourse Analysis and Theoretical Lenses

To investigate the discourse on frameworks, principles, and practices of AI fairness, we employ a research approach that combines discourse analysis with thematic analysis through a praxiological socio-technical lens. The study examines discourse, i.e., "all forms of talk and texts" (Gill, 2000, p. 174). We focus on the content of AI fairness governance rather than its discourse process, and the discourse analysis serves as a perspective, not a method (van Dijk, 2009), guiding the design of the practices.

We adopt Cukier et al.'s (2009) four-step methodology, combining Richardson's (2007) newspaper analysis and Gill's (2000) approach. The corpus definition is identified as the first step through the application of exclusion criteria. Steps 2, 3, and 4 of content analysis, observation interpretation, and findings explanation utilized thematic analysis.

## 3.1 Corpus Definition

Our scope of discourse is twofold: discourse and AI Fairness governance. First, we base our inclusion criteria on the definition of discourse as presented in various publications. The object is publishers' communication in newspapers, magazines, and academic manuscripts, including both academic and grey literature, to represent non-academic public discourse. The second scope criterion is AI fairness governance, encompassing guidelines, principles, processes, roles, and strategies.

The wide variety of object types and sources enables the most inclusive consideration of the relevant discourse. Table 1 depicts the various data sources, their inclusion criteria, sources, and search details.

| Object Type | Inclusion | Database/ Source | Initial Search | Excluded | Forward/ Backward | Corpus items |
|---|---|---|---|---|---|---|
| Newspapers | Geography: UK, US<br>Area: Business<br>Readers: Top 10 | Factiva | 111 | 109 | 0 (42) | 2 |
| Magazines | Geography: UK, US<br>Area: Business<br>Readers: Top 50 | Business Source Complete | 50 | 38 | 4 (1,336) | 16 |
| Reports | Government, management consultancies, experts, European Commission, experts | Bing search | 314 | 288 | 9 (2,462) | 35 |
| Academic manuscripts | FT50<br>AIS11 | Web of Science | 8 | 3 | 2 (833) | 7 |
| SUM | | | 483 | 438 | 15 | 60 |

*Table 1.     Data Corpus and Search Details.*

The unit of **analysis** is an object, i.e., a publication of a newspaper article, an academic manuscript, or an interview. Each object is analyzed for two content types: what and how it is communicated. All data corpus articles are incorporated in their original form and not summarized or corrected (Gill, 2000).

We combined through Boolean AND the terms related to AI fairness ("AI Fairness" or "AI justice"), governance ("governance" or "management"), and frameworks ("framework" or" guideline" or "principle" or guidance"). The databases searched are presented in Table 1 and are adapted to the object type. The initial search resulted in 483 sources. We then applied the exclusion criteria of duplicate or near-duplicate (95% same text), references to other AI fairness publications only, no details on AI fairness governance provided, and no specific guidelines or details on AI fairness governance given. For example for newspapers, initially 111 were included as they fulfiled the search terms, but were then excluded mainly based the exclusion criteria three "No details on AI fairness governance or only examples of AI fairness are given" (69%) and four "No specific guidelines or details on AI fairness governance given" (47%), where multiple criteria in parallel possible. This resulted in 45 objects to include. An unrestricted forward and backward search reviewed an additional 4,673 objects, adding 15 objects. The final sample[1] consisted of 60 objects: Two newspaper articles, 16 magazine articles, 35 reports, and seven manuscripts.

## 3.2  Content Analysis, Observation Interpretation, and Findings Explanation

The coding of the articles and their analysis to identify observations of validity claims and elements of the AI fairness governance frameworks is part of the thematic analysis. Thematic analysis identified, analyzed, and reported the themes that defined the AI fairness governance. Aligned with Braun and Clarke (2006), our theme, governing AI for fairness, represents "something important about the data" (Braun & Clarke, 2006, p. 82), grounded in the data's importance and prevalence. Secondly, the thematic analysis focuses on a detailed account of AI fairness governance. The sub-themes identified within AI fairness governance are inductively identified and linked to the data (Braun & Clarke, 2006), representing a data-driven thematic analysis (Braun & Clarke, 2006). The data collection and analysis focus solely on AI fairness governance, representing a semantic or explicit level of thematic analysis (Braun & Clarke, 2006). Aligned with the socio-technical lens, the thematic analysis is grounded in a constructionist epistemology, enabling theorizing within the sociocultural context (Braun & Clarke, 2006).

We followed the six-step approach to thematic analysis (Braun & Clarke, 2006), which aligns with the research paradigm of this study, given its ontological and epistemological flexibility (Braun & Clarke, 2006). The phases of thematic analysis (Braun & Clarke, 2006) are presented in Table 2.

| Phase | Activities |
| --- | --- |
| 1. Familiarize with Data | We uploaded all data corpus objects to MAXQDA and read them multiple times. A project memo in MAXQDA captured our early coding ideas and initial patterns, enabling a critical stance toward the data and suspending taken-for-granted assumptions. |
| 2. Generate Initial Codes | We developed an inductive a priori coding scheme, informed by the literature and praxiologic and socio-technical lenses. The codes were then applied, adapted, and extended in MAXQDA. The coding was inclusive and comprehensive and was discussed between authors one and two in multiple workshops to achieve coding congruence. |
| 3. Search for Themes & Categories | We formed themes (such as AI fairness practices) and categories (such as AI fairness goals) through code collation, hierarchical ordering, and relational connection. Frequency tools and code maps in MAXQDA were utilized to derive the initial thematic map, which was documented in detail in a MAXQDA memo. We refined the thematic map iteratively after each data corpus object was coded. |
| 4. Review Themes & Categories | We evaluated the candidate themes and categories for internal homogeneity and external heterogeneity, and added further details, including sub-categories and relationships among all parts of the thematic map. The thematic map was further updated with corrections and additions based on an additional check against the complete data set, using the MAXQDA code matrix browser and smart coding tool. |
| 5. Define and Name Themes | We then named and described the final themes, categories, and sub-categories, including definitions, narrative synopses, and relational mappings. These were documented in the updated thematic map to reflect analytical clarity and thematic coherence. |

---

[1] The details of the final data corpus can be accessed on
https://osf.io/r732y/overview?view_only=5dceb78992c04312bce4890195b01977.

| | |
|---|---|
| 6. Produce Report | We documented the final thematic map description in this manuscript and integrated verbatim transcript extracts from the data corpus. We also linked the themes to the research question of how organizations implement AI fairness, i.e., the proposed practices, and situated them within the extant literature. |

*Table 2.        Thematic Analysis Phases and Implemented Activities.*

Connections between sub-categories and sub-categories to categories were established based on associations between codes, and then hierarchically ordered as part of step three in the thematic analysis. For example, the code "Train users/employees" describes the STS element of a task part of the sub-category "User Training", as shown in the comment of "Training should be delivered to AIDA system users, which could cover topics including understanding of system process, confidence level, and uncertainties of outcomes, override provision and potential likely unfairness scenarios (MAS et al., 2022, p. 43).

### 3.3    Praxiological and Socio-Technical Lenses

We adopt praxiological and socio-technical lenses to develop practice-focused rather than principles-based findings. While principles can guide action (Whittlestone et al., 2019) and inform decisions (Chatterjee et al., 2024), their proliferation (Floridi & Cowls, 2021; Morley et al., 2023) has not closed the translation gap, i.e., the failure to convert principles into practices (Ali et al., 2023; Mittelstadt, 2019). This gap stems from vague, practitioner-inadequate, and non-actionable guidance (Baldassarre et al., 2024; Morley et al., 2023). Principles are implemented through practices (Seppälä et al., 2021). We propose AI fairness practices to bridge this gap (Horneber, 2025). Praxeology (Nicolini, 2017) states that activities unfold through materially mediated practices (Schatzki, 2001). Practices are "regimes of a mediated object-oriented performance of an organized set of sayings and doings" (Nicolini, 2017, p. 5), representing technology as material objects, as social relations between agents and technology, and as institutional properties (Nicolini et al., 2021). AI fairness practices are useful and actionable for AI practitioners, enabling them to achieve and maintain AI fairness through their concreteness and operationalization. They are normative, as they represent norms that guide expected behavior, aiming to change behavior through compliance with these practices and to achieve AI fairness. AI fairness practices integrate into the wider organizational governance and target internal organizational agents. Effective practices are evidence-based (grounded in empirical data), context-sensitive (adapted to organizational context), legitimate (socially accepted), embedded (integrated with other governance and practices), inclusive (includes diverse stakeholder views), and aligned with the goal of AI fairness. Each practice description is influenced first by the context-mechanism-outcome (CMO) structure (Pawson & Tilley, 1997), in which mechanisms are further delineated into the socio-technical components of task, agent, technology, and structure. We ensured that the three components of ThinkLets (thinking patterns (Briggs & De Vreede, 2009) of collaboration engineering (Briggs et al., 2006), of tools, configuration, and scripts (Briggs & De Vreede, 2009; Schwabe et al., 2025) are covered.

STS serves as our lens for examining how AI fairness practices are structured inclusively, addressing gaps such as the overemphasis on technical guidance and the underrepresentation of context and agent influence on AI fairness. Originating with Trist and Bamforth's (1951) study, STS has, over time, become a cohesive axis of interest for information systems (IS) research (Sarker et al., 2019), conceptualized as systems comprising technical and social subsystems. The technical subsystem includes technical and material artifacts, and the techniques or practices employed to use the artifacts (Lee, 2004), while the social component encompasses individuals or collectives and the relationships between them, which might be expressed as roles, hierarchies, structures, economic systems, cultures, power relations, or communication networks (Sarker et al., 2019).

*Short Title (up to 5 words)*

## 4 Results

### 4.1 Overall Findings

We developed a thematic map as a visual and conceptual representation of the analysis findings, and it structures the emergent themes into a coherent framework. At the top level, the map delineates major themes that encapsulate the AI fairness framework, which are further broken down into categories that reflect specific areas of focus, such as defining AI fairness and enumerating the AI fairness practices. AI Governance frameworks ensure the implementation of AI Fairness practices, which lead to AI Fairness. Each category is supported by conceptual dimensions that articulate the underlying principles and tensions observed in the literature. Figure 1 depicts the thematic map of the AI fairness framework.

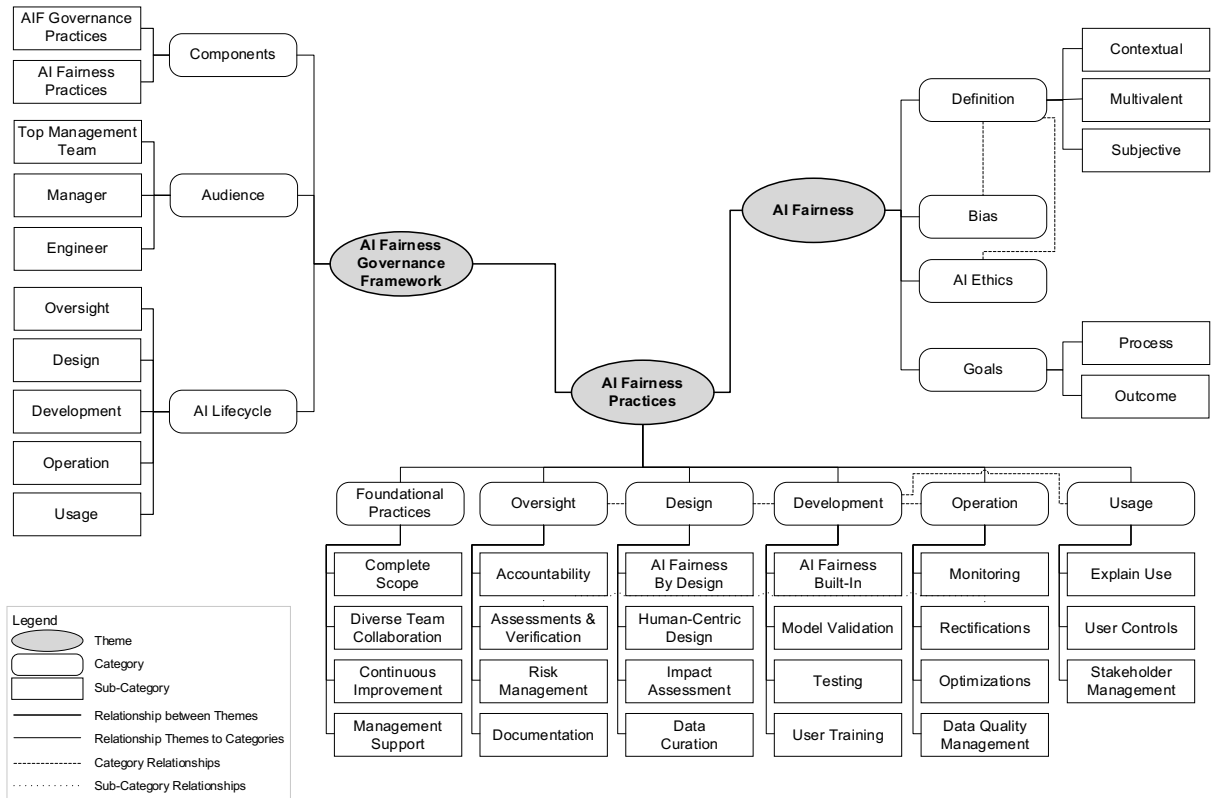

*Figure 1.        AI Fairness Governance Thematic Map.*

Various interactions were identified and depicted as relationships. The theme of AI Fairness Practices encompasses practices for each lifecycle phase and foundational practices (i.e., sub-categories). For example, the identified AI fairness definition is related to the category AI Ethics, as AI fairness forms one part of AI ethics. Describing all elements of the thematic analysis is beyond the scope of this paper. Still, we found the AI fairness definition, the components of an AI fairness governance framework, and actionable AI fairness practices most interesting.

**AI fairness**. We identified multiple *definitions* of AI fairness in the analyzed data corpus. Some define it with a focus on data, such as "data are used in a manner most useful to stakeholders" (Abrams et al., 2019, p. 22), and others with a focus on bias, such as "AI systems built without bias" (Asif et al., 2023, p. 3). Based on the synthesis and integration of definitions, we refer to AI fairness as "the context-dependent, multivalent, and actor-relative judgment that AI behaves in ways that are equal, just, and reasonable, ensuring non-discrimination and equality". AI fairness is context-dependent, as it manifests differently across environments (Leslie et al., 2023), and multivalent, i.e., it has different meanings depending on the subject to which it relates, such as data, models, or systems. It is also subjective, as it is a social judgment by an individual and therefore yields different outcomes. The issue of *bias*, i.e., the "tendency, inclination, or prejudice toward or against something or someone" (Smith & Rustagi, 2020,

p. 20), is addressed through AI fairness, a component of *AI ethics* (Jobin et al., 2019). AI fairness entails balancing the competing *goals* of distributive non-discrimination and procedural equality, thereby ensuring both equitable outcomes and just processes in the design, development, and use of AI.

**AI fairness governance framework**. We identified across the data corpus that, for maximum utility, a governance framework needs to include not only practices themselves as *component*, but also practices for governance. The AI Fairness Governance Practices represent the why and what, or practices to govern for AI fairness, while the AI Fairness Practices represent the how or practices to implement and manage AI fairness. Based on the synthesized literature, we define AI fairness governance as the system of rules, practices, processes, and technological tools that are employed to ensure an organization's use of AI technologies aligns with the organization's strategies, objectives, and values; fulfills legal requirements; and meets principles of fair AI followed by the organization. The socio-technical framework enables organizations to balance achieving AI Fairness with leveraging IT to achieve organizational objectives while minimizing risks and complying with relevant laws and regulations across the AI lifecycle. Our analysis of the data revealed that organizations require AI fairness governance practices to achieve AI fairness, ranging from defining and agreeing on the overarching objectives for AI fairness to allocating decision rights to determine which roles make which decisions on AI fairness, including which practices to implement. Furthermore, AI fairness governance practices should be tailored to the organization's context and train salient stakeholders in AI fairness generally, as well as in specific practices to follow. Different roles in the organization, based on their role definitions, require different guidance to perform their duties effectively. We distinguish three roles in the *audience* for AI Fairness practices: Top Management Team, Managers, and Engineers. The evidence supported extending the traditional *AI lifecycle* with an oversight phase to orchestrate accountabilities, manage risks, and assess AI fairness. The design phase includes discover, design, and curate activities, where the development phase builds, trains, tests, deploys, and trains users. The operation phase sustains, maintains, monitors, and retires, while the usage phase explains use and manages stakeholders.

**AI Fairness practices.** Based on the data, we identified foundational or lifecycle category practices. Foundational AI fairness practices apply to all lifecycle phases, focusing on how to behave and implement AI fairness. For example, the data clearly demonstrated that management support, ensuring leadership prioritization through endorsement and resource allocation, is critical for all phases of the AI lifecycle to be effective. Lifecycle-specific AI fairness practices guide organizations in practically implementing AI fairness, as described in the AI Fairness Practice Matrix (see the next section).

Our analysis of the dataset reveals that prevailing guidance on AI fairness remains predominantly general and lacks specificity (Cachat-Rosset & Klarsfeld, 2023). This contributes to the persistent "principles-to-practice" quandary (McGrath et al., 2024, p. 10), wherein organizations struggle to translate high-level recommendations into actionable steps (Benjamins et al., 2019). The findings underscore the need for operationalized, detailed practices that concretely support the implementation of AI fairness as an objective. Notably, most existing guidance focuses on the organizational level, often overlooking the need for role-specific instructions. Role-tailored guidance facilitates more effective operationalization by delineating clear responsibilities and recommending appropriate tools for particular tasks. Furthermore, the analysis of the dataset indicates that the currently provided guidance is frequently limited to discrete phases of the AI lifecycle, most prominently the design phase, without offering an integrated, end-to-end perspective that spans, governance, design, development, deployment, operation, and usage. Comprehensive guidance must encompass the entire AI lifecycle and address interdependencies between lifecycle phases and associated practices.

Next, we present only one AI fairness practice in the next section, as presenting all would exceed the space requirements. All practices are provided in the supplementary online materials.

### 4.2  AI Fairness Practices

Synthesizing the findings leads to the development of governance practices for AI fairness. The **AI Fairness Practice Matrix** structures practices by AI lifecycle phase and uses consecutive numbering, titles, and a standard structure-based description. Each practice Practice obligation degree: Not all practices are equally important in achieving AI fairness. To provide nuanced guidance, we introduce the obligation degrees for each practice: "Must" represents a requirement and compliance is mandatory, "Should" is a recommendation and it is advised to implement but not mandatory, "Could" represents a suggestion and is optional, and "May" is an option, offering discretion. Each practice is categorized by the audience for whom it is relevant. The AI lifecycle-based phases are used to define which lifecycle phase each practice applies to.

We present the four foundational and 19 AI Fairness practices in a matrix[2]. To provide insights into the derived detailed practices, we present Table 3, the development-phase practice of "Testing". These are based on the data corpus objects and link directly back to identified guidance. Examples include "companies implementing Responsible AI should rigorously test their models for not only safety and fairness but also accuracy" (Jones Suzuki & Namireddy, 2024, p. 4) and "extensive internal testing to validate the effectiveness of any AI" (King, 2024, p. 2). Each component and its definition of an AI fairness practice are based on the codes allocated in the data corpus. They include the STS elements context, task, tool, and actor, which are interdependent as per STS, but were amended with other identified components, such as subsystems and their connections or metrics, to result in a comprehensive and operationalised practice. The praxiological lens drove the action-oriented components and the overall definition of all AI fairness practices. The combination of all practice components (rows in Table 3) ensures concreteness and operationalization of AI fairness practices. Each effective practice is intended to be evidence-based (grounded in empirical data), context-sensitive, legitimate, embedded, inclusive, and aligned with the goal of AI fairness. The structure of practice definitions is influenced first by the context-mechanism-outcome (CMO) structure (Pawson & Tilley, 1997), in which mechanisms are further delineated into the socio-technical components of task, agent, technology, and structure. The operationalized AI fairness practices represent the "doing of fairness" as they implement activities to achieve and maintain AI fairness. The thematic map theme of 'AI Fairness Practices' in Figure 1 is enumerated and further detailed in the structure in Table 3. This table not only details, based on the coding, hierarchical ordering, and relational connections, the AI fairness practices, but especially specifies the STS elements.

| Phase | Development | |
|---|---|---|
| Name | 4.3 Testing | |
| Definition | Outcome | The Testing practice validates AI and ensures all system components are fair and compliant. |
| | Subsystems | It integrates adversarial testing, compliance audits, performance benchmarking, and simulation-based evaluations, with cross-functional collaboration and governance oversight mechanisms, to achieve robustness, fairness, safety, and regulatory compliance of AI systems prior to and throughout deployment. |
| | Context | It operates within organizational and regulatory contexts shaped by sector-specific standards, legal obligations, and fairness principles governing AI. |
| | Tasks | Core activities include specifying functional and non-functional test objectives (e.g., AI fairness, bias, robustness, explainability) and operationalized criteria; designing test scenarios and protocols; executing tests to verify that AI performs as intended (e.g., adversarial testing); documenting test results; and rectifying identified AI fairness issues. |
| | Tools | These activities are supported by tools, e.g., for robustness checks and AI fairness, as well as regulatory sandboxes. |
| | Actors | The practice assigns responsibilities across actors such as AI engineers, compliance officers, fairness review boards, and external auditors, ensuring |

---

[2] The details of AI Fairness Practice Matrix can be accessed on
https://osf.io/r732y/overview?view_only=5dceb78992c04312bce4890195b01977.

| | | |
|---|---|---|
| | | that AI systems meet defined performance thresholds, mitigate harmful outputs, and uphold procedural fairness across demographic subgroups. |
| | Metrics | Effectiveness is assessed through metrics that capture model accuracy, robustness, fairness, and explainability, while emergent challenges arise from performance drift, hidden proxies of bias, and the evolving nature of adversarial threats and regulatory expectations. |
| | Success Factors | Critical success factors include interdisciplinary testing protocols, transparent documentation, and continuous evaluation aligned with Responsible AI principles. |
| Obligation Degree | Must | |
| Interrelations | Input to 5.2 Rectifications | |
| Audience | ☐ Top Management Team     ☐ Manager     ☒ Engineer | |
| Applicability | Development - Test | |

*Table 3.      AI Fairness Practice Matrix Example Testing Practice.*

# 5      Discussion

This study examined socio-technical practices for AI fairness and operationalized AI fairness through actionable practices that integrate technical and social dimensions. The thematic map illustrates the interconnected nature of AI fairness and provides details on its context-dependence, multivalence, and subjectivity. It balances competing objectives of distributive non-discrimination and procedural equality. The AI Fairness Practice Matrix addresses the fragmented guidance gap by proposing a holistic, integrated set of AI fairness practices spanning the entire AI lifecycle. It provides four foundational and 19 AI fairness practices, graded by obligation degree (must/should/could/may), and mapped to organizational roles (top management/managers/engineers).

The contribution of this research is to extend the AI fairness problematization beyond technical compliance to encompass socio-technical integration and to address the five identified limitations of existing guidance. We provide focused guidance on implementing AI fairness, not generic AI ethics. The proposed AI fairness practices provide operationalized, actionable guidance that closes the principle-to-practice gap (Ali et al., 2023), enabling organizations to adapt and implement efficient, effective AI fairness practices. The developed AI fairness practices are consistently structured and detailed, utilizing socio-technical systems and ThinkLets inputs to ensure completeness for the level of detail and full AI lifecycle coverage. Additionally, the practices cover the entire AI lifecycle in an integrated way, leading to effective practices in achieving and maintaining AI fairness. The AI fairness practice matrix and its practices are adaptable to changing organizational contexts, providing the required dynamism and applicability to implement and sustain AI fairness effectively.

## 5.1      Implications for Research

The paper contributes to a practice turn within a STS account of AI fairness (Baldassarre et al., 2024; Bateni et al., 2022; Lu et al., 2024; Lu et al., 2023; Mäntymäki et al., 2022). It does not stop at abstract values or metric trade-offs but goes beyond that, providing specific, operational guidance, based on multiple sources, described as structured practices. By systematizing the description of the practices, it delivers building blocks that can be combined to govern AI fairness more effectively. The fact that the practices emerge as parts of a larger framework rather than a loosely coupled set (cf. Lu et al. (2023), Lu et al. (2024)) shows that the combination of practices needs to be carefully considered. Whereas the proposed practices provide a first intuition for how they could be combined by indicating their interrelatedness or type, the study of interdependencies should go further, potentially leading to the discovery of chains that can be transferred across contexts. The modular nature of practices presented here allows for their easy recombination and testing. Their grounding in circulating frameworks from research and industry allows them to reference existing guidance and potential experience reports that may emerge over time, whether the practitioner draws on the practices presented in this paper or from the guidance in one of our sources. This opens further avenues for research in this area that could incorporate experiences concerning the effectiveness of the practices.

The manuscript also strengthens the discourse on specific features of AI fairness, rather than the generic focus on responsible/trustworthy AI (Baldassarre et al., 2024; Barletta et al., 2023). Especially, the analysis confirms that AI fairness is frequently presented as multivalent, subjective, and contextual. Accordingly, it demands a sociotechnical treatment, inherent to information systems rather than a purely technical and distributive account. Moving further, it opens new avenues of research with particular relevance to IS. Exploring how the ambiguity of fairness interacts with organizational goals and the clear measures that might be necessary for legal reasons could yield new insights into the interplay among technology, organizations, and human values.

Finally, the manuscript offers a new format for describing governance practices, which, on the one hand, relies on an acknowledged script format influenced by CMO, STS, and ThinkLets structure (Briggs & De Vreede, 2009; Pawson & Tilley, 1997; Trist & Bamforth, 1951) while also incorporating organizational and processual aspects, such as the audience and applicability scope. Given that governance guidance frequently remains vague in terms of who should apply it and when (Floridi & Cowls, 2021), the combination of relatively specific scripts with those aspects provides essential advantages for companies that want to employ the practices.

The graded obligations (Must/Should/Could/May) provide mechanisms for adaptive, context-sensitive governance responsive to evolving stakeholder expectations and socio-technical conditions in use, addressing the negotiable and ambiguous nature of fairness (Chowdhury & Klautzer, 2025; Dolata & Schwabe, 2023). This addresses the documented required dynamic treatment of fairness in many frameworks (Dolata & Schwabe, 2023).

## 5.2 Implications for Practice

For practitioners, the AI Fairness Practice Matrix provides role clarity and timing by articulating who does what, when, and to what standard, making accountability auditable and regulator-ready without overburdening engineering teams with abstract principle lists. The audience-phase mapping translates justice dimensions into concrete activities for different audiences across design, development, operation, use, and oversight, generating traceable artifacts (e.g., thresholds, documentation, training records) that support internal reviews and external assurance. This directly addresses the shortcomings across frameworks—missing role specificity, partial lifecycle coverage, and static guidance —by offering a navigable governance scaffold embedded within the existing AI lifecycle.

Operationally, the minimal set of "Must" practices yields an immediately implementable fairness control baseline. This could be composed of, e.g., clear accountability with decision rights and regular AI fairness assessments in oversight, AI fairness designed into the system from the start, and fair data curated in design, validated models that are robustly tested with data in development, continuous monitoring drives optimizations for AI fairness in operations, and comprehensible explanations of AI in usage. Organizations can leverage pattern-oriented process engineering and lifecycle governance patterns to standardize and scale these approaches, enabling consistent execution across teams and products. The components of a practice and across all AI fairness practices require fit, i.e., the match between them (Venkatraman, 1989). The proposed AI fairness practices may not constitute neutral additions to existing practices, and the obligation degree is context-specific. For instance, practices graded as having an obligation level of "Must" for engineers (e.g., extensive AI fairness testing) can create workload and prioritisation conflicts for managers tasked with delivering projects under tight deadlines. This tension highlights trade-offs between control and flexibility, and, in our context, AI fairness practices reconfigure them (Murire, 2024) by introducing new accountability demands within AI governance (Birkstedt et al., 2023). This aligns with institutional logic (Kostova & Zaheer, 1999), which guides behavior through practices (Thornton & Ocasio, 1999, p. 804). The different identified audiences of the top management team, the manager, and the engineer could perceive tensions between role-based goals and AI fairness practices.

Yet, most importantly, the presented practices guide practitioners overwhelmed by the sheer number of guides and frameworks for assuring AI fairness by integrating them into a logically structured set of practices. This provides support for navigating and sorting the available scientific and practitioner articles. In complex, overloaded topics like AI fairness with profound ethical, regulatory, and practical implications, such guidance-for-guidance can offer an essential step towards implementing the necessary AI fairness measures.

## 5.3    Limitations

First, this study centers on AI fairness as a single ethical principle, though fairness interacts with other principles (e.g., transparency or accountability) and non-ethical considerations (e.g., organizational performance). While this focus enables depth, it limits the analysis of interdependencies. Future research should extend the practices to incorporate multiple ethical dimensions and empirically validate their combined governance impact. Second, findings are interpretive, grounded in the published discourse from academic, policy, and industry sources. To improve transferability, future work should complement this with empirical studies, such as organizational case studies and cross-sectoral surveys, to capture lived practices and validate proposed practices. Third, identified practices are context-dependent and subject to change as regulations, organizational priorities, and technologies evolve. Emerging paradigms, such as agentic and generative AI, introduce dynamics in human–machine relationships that existing practices may not fully address (Laine et al., 2025). Future research should explore adaptive governance mechanisms responsive to these developments. Finally, the practice matrix considers three stakeholder groups, potentially oversimplifying the fairness complexity. A more nuanced understanding of stakeholder diversity is needed to address "fair to whom?". Future studies should therefore refine and expand the matrix to account for additional stakeholder views and context-specific fairness requirements.

## 5.4    Conclusion

In this study, we advance the discourse on AI fairness by proposing structured, actionable practices that address the principles-to-practice gap. By constructing the AI fairness practice matrix, comprising four foundational and 19 lifecycle-specific practices, the research operationalises AI fairness as a context-dependent, multivalent, and socially constructed concept. In contrast to existing guidance, we offer a modular, dynamic set of practices to guide different roles in organizations in implementing and maintaining AI fairness. We contribute to the IS literature by extending AI fairness beyond abstract ethical principles toward concrete, role-specific interventions. While we grounded the practices in extensive discourse and thematic analysis, future research should empirically validate their efficacy and inter-practice dependencies.

# References

Abrams, M., Abrams, J., Cullen, P., & Goldstein, L. (2019). Artificial intelligence, ethics, and enhanced data stewardship. *IEEE Security & Privacy*, *17*(2), 17-30.

Ali, S. J., Christin, A., Smart, A., & Katila, R. (2023). Walking the walk of AI ethics: Organizational challenges and the individualization of risk among ethics entrepreneurs. ACM Conference on Fairness, Accountability, and Transparency (FAccT '23), Chicago, US.

Asif, R., Hassan, S. R., & Parr, G. (2023). Integrating a Blockchain-Based Governance Framework for Responsible AI. *Future Internet*, *15*(3), 97.

Baldassarre, M. T., Gigante, D., Kalinowski, M., & Ragone, A. (2024). POLARIS: A framework to guide the development of trustworthy AI systems. Proceedings of the IEEE/ACM 3rd International Conference on AI Engineering-Software Engineering for AI,

Bargh, M. S., Choenni, S., & Ter Braak, F. (2025). Algorithmic fairness as sociotechnical system. 26th Annual International Conference on Digital Government Research, Porto Alegre City, BRA.

Barletta, V. S., Caivano, D., Gigante, D., & Ragone, A. (2023). A rapid review of responsible AI frameworks: How to guide the development of ethical AI. Proceedings of the 27th International Conference on Evaluation and Assessment in Software Engineering, Oulu, FIN.

Bateni, A., Chan, M. C., & Eitel-Porter, R. (2022). AI fairness: from principles to practice. *arXiv preprint*, *arXiv:2207.09833*, 1-21.

Benjamins, R., Barbado, A., & Sierra, D. (2019, 7-9/11/2019). Responsible AI by design in practice. Proceedings of the Human-Centered AI: Trustworthiness of AI Models & Data (HAI), Washington, US.

Binns, R. (2018). Fairness in machine learning: Lessons from political philosophy. Conference on fairness, accountability and transparency, New York, US.

Birkstedt, T., Minkkinen, M., Tandon, A., & Mäntymäki, M. (2023). AI governance: themes, knowledge gaps and future agendas. *Internet Research*, *33*(7), 133-167.

Bostrom, R. P., & Heinen, J. S. (1977). MIS problems and failures: A socio-technical perspective Part I: The causes. *MIS Quarterly*, *1*(3), 17-32.

Bradfield, M., & Aquino, K. (1999). The effects of blame attributions and offender likableness on forgiveness and revenge in the workplace. *Journal of Management*, *25*(5), 607-631.

Braun, V., & Clarke, V. (2006). Using thematic analysis in psychology. *Qualitative research in psychology*, *3*(2), 77-101.

Briggs, R., Kolfschoten, G., Vreede, G.-J., & Douglas, D. (2006). Defining key concepts for collaboration engineering. Proceedings of Americas Conference on Information Systems (AMCIS), Acapulco, MEX.

Briggs, R. O., & De Vreede, G.-J. (2009). *ThinkLets: building blocks for concerted collaboration*. University of Nebraska

Butcher, J., & Beridze, I. (2019). What is the state of Artificial Intelligence governance globally? *The RUSI Journal*, *164*(5-6), 88-96.

Cachat-Rosset, G., & Klarsfeld, A. (2023). Diversity, Equity, and Inclusion in Artificial Intelligence: An evaluation of guidelines. *Applied Artificial Intelligence*, *37*(1), 1-29.

Cadbury, A. (1992). *Report of the committee on the financial aspects of corporate governance* (0852589131). G. a. d. o. P. P. Ltd.).

Chowdhury, S., & Klautzer, L. (2025). Shaping an adaptive approach to address the ambiguity of fairness in AI: Theory, framework, and illustrations. *Cambridge Forum on AI: Law and Governance*, *1*.

Christian, B. (2021). *The alignment problem: How can machines learn human values?* Atlantic Books

Colquitt, J. A. (2012). Organizational justice. In S. W. J. Kozlowski (Ed.), *The Oxford Handbook of Organizational Psychology* (pp. 526-547). Oxford University Press

Colquitt, J. A., Greenberg, J., & Zapata-Phelan, C. P. (2005). What is organizational justice? A historical overview. In J. Greenberg & J. A. Colquitt (Eds.), *Handbook of Organizational Justice* (pp. 3-58). Lawrence Earlbaum Associates

Cukier, W., Ngwenyama, O., Bauer, R., & Middleton, C. (2009). A critical analysis of media discourse on information technology: preliminary results of a proposed method for critical discourse analysis. *Information Systems Journal*, *19*(2), 175-196.

Davenport, T. H., & Redman, T. C. (2025). How to marry process management and AI. *Harvard Business Review*, *January-February*.

De Almeida, P. G. R., Dos Santos, C. D., & Farias, J. S. (2021). Artificial Intelligence Regulation: a framework for governance. *Ethics and Information Technology*, *23*(3), 505-525.

Decker, M. C., Wegner, L., & Leicht-Scholten, C. (2025). Procedural fairness in algorithmic decision-making: the role of public engagement. *Ethics and Information Technology*, *27*(1).

Dignum, V. (2018). Ethics in artificial intelligence: introduction to the special issue. *Ethics and Information Technology*, *20*(1), 1-3.

Dolata, M., Feuerriegel, S., & Schwabe, G. (2022). A sociotechnical view of algorithmic fairness. *Information Systems Journal*, *32*(4), 754-818.

Dolata, M., & Schwabe, G. (2023). Towards the socio-algorithmic construction of fairness: The case of automatic price-surging in ride-hailing. *International Journal of Human–Computer Interaction*, 1-11.

Feuerriegel, S., Dolata, M., & Schwabe, G. (2020). Fair AI. *Business & Information Systems Engineering*, *62*(4), 379-384.

Floridi, L., & Cowls, J. (2021). A unified framework of five principles for AI in society. In L. Floridi (Ed.), *Ethics, Governance, and Policies in Artificial Intelligence* (pp. 5-18). Springer

Gill, R. (2000). Discourse Analysis. In M. W. Bauer & G. Gaskell (Eds.), *Qualitative Researching with Text, Image and Sound* (Vol. 1, pp. 172-190). SAGE Publications

Greenberg, J. (1986). Determinants of perceived fairness of performance evaluations. *Journal of Applied Psychology*, *71*(2), 340-342.

Greenberg, J. (1990). Organizational justice: Yesterday, today, and tomorrow. *Journal of Management*, *16*(2), 399-432.

Greenberg, J. (1993). The social side of fairness: Interpersonal and informational classes of organizational justice. In R. Cropanzano (Ed.), *Justice in the Workplace: Approaching Fairness in Human Resource Management* (pp. 79-106). Lawrence Erlbaum Associates

Horneber, D. (2025). Understanding the implementation of responsible Artificial Intelligence in organizations: A neo-institutional theory perspective. *Communications of the Association for Information Systems*, *57*(1), 8.

Jobin, A., Ienca, M., & Vayena, E. (2019). The global landscape of AI ethics guidelines. *Nature Machine Intelligence*, *1*(9), 389-399.

Jones Suzuki, G., & Namireddy, M. (2024). *Guidance for Implementing Responsible AI in Legal and Business Practice*. R. Gray. https://www.ropesgray.com/en/insights/alerts/2024/05/guidance-for-implementing-responsible-ai-in-legal-and-business-practice

Kals, E., & Jiranek, P. (2012). Organizational justice. In E. Kals & J. Maes (Eds.), *Justice and Conflicts: Theoretical and Empirical Contributions* (pp. 219-235). Springer

King, J. (2024, 29/01/2024). BankThink Banks cannot skimp on AI vendor vetting. *American Banker*, 1-3.

Kostova, T., & Zaheer, S. (1999). Organizational legitimacy under conditions of complexity: The case of the multinational enterprise. *Academy of Management Review*, *24*(1), 64-81.

Laine, J., Minkkinen, M., & Mäntymäki, M. (2025). Understanding the ethics of generative AI: Established and new ethical principles. *Communications of the Association for Information Systems*, *56*, 1-25.

Lee, A. S. (2004). Thinking about social theory and philosophy for information systems. In J. Mingers & L. Willcocks (Eds.), *Social theory and philosophy for information systems* (Vol. 1, pp. 1-26). John Wiley & Sons

Leslie, D., Rincón, C., Briggs, M., Perini, A., Jayadeva, S., Borda, A., Bennett, S., Burr, C., Aitken, M., Katell, M., Fischer, C., Wong, J., & Kherroubi Garcia, I. (2023). *AI Fairness in Practice*.

Li, B., Qi, P., Liu, B., Di, S., Liu, J., Pei, J., Yi, J., & Zhou, B. (2023). Trustworthy AI: From principles to practices. *ACM Computing Surveys*, *55*(9), 1-46.

Li, H., Sarathy, R., Zhang, J., & Luo, X. (2014). Exploring the effects of organizational justice, personal ethics and sanction on internet use policy compliance. *Information Systems Journal*, *24*(6), 479-502.

Lu, Q., Zhu, L., Whittle, J., & Xu, X. (2024). *Responsible AI: Best practices for creating trustworthy AI systems*. Addison-Wesley Professional

Lu, Q., Zhu, L., Xu, X., Whittle, J., Zowghi, D., & Jacquet, A. (2023). Operationalizing responsible AI at scale: CSIRO Data61's pattern-oriented responsible AI engineering approach. *Communications of the ACM*, *66*(7), 64-66.

Mäntymäki, M., Minkkinen, M., Birkstedt, T., & Viljanen, M. (2022). Defining organizational AI governance. *AI and Ethics*, *2*(4), 603-609.

Mäntymäki, M., Minkkinen, M., Zimmer, M. P., Birkstedt, T., & Viljanen, M. (2023). Designing an AI governance framework: From research-based premises to meta-requirements. Thirty-first European Conference on Information Systems (ECIS 2023), Kristiansand, Norway.

Marcinkowski, F., Kieslich, K., Starke, C., & Lünich, M. (2020, 2020). Implications of AI (un-)fairness in higher education admissions. Proceedings of the 2020 Conference on Fairness, Accountability, and Transparency, Barcelona, ESP.

MAS, Accenture, & SwissRe. (2022). *Veritas Document 3A: FEAT Fairness Principles Assessment Methodology*. https://www.mas.gov.sg/-/media/mas-media-library/news/media-releases/2022/veritas-document-3a---feat-fairness-principles-assessment-methodology.pdf

McGrath, Q., Hevner, A. R., & Vreede, G.-J. D. (2024). *Managing ethical risks of Artificial Intelligence in business applications*. Institute of Electrical and Electronics Engineers (IEEE). https://dx.doi.org/10.36227/techrxiv.170905835.50964792/v1

Mitchell, S., Potash, E., Barocas, S., D'Amour, A., & Lum, K. (2021). Algorithmic fairness: Choices, assumptions, and definitions. *Annual Review of Statistics and Its Application*, *8*(1), 141-163.

Mittelstadt, B. (2019). Principles alone cannot guarantee ethical AI. *Nature Machine Intelligence*, *1*(11), 501-507.

Morley, J., Kinsey, L., Elhalal, A., Garcia, F., Ziosi, M., & Floridi, L. (2023). Operationalising AI ethics: Barriers, enablers and next steps. *AI & SOCIETY*, *38*(1), 411-423.

Murire, O. T. (2024). Artificial Intelligence and its role in shaping organizational work practices and culture. *Administrative Sciences*, *14*(12), 1-16.

Nicolini, D. (2017). Practice theory as a package of theory, method and vocabulary: Affordances and limitations. In M. Jonas, B. Littig, & A. Wroblewski (Eds.), *Methodological reflections on practice oriented theories* (pp. 19-34). Springer International Publishing

Ochmann, J., Michels, L., Tiefenbeck, V., Maier, C., & Laumer, S. (2024). Perceived algorithmic fairness: An empirical study of transparency and anthropomorphism in algorithmic recruiting. *Information Systems Journal*, n/a-n/a.

Pawson, R., & Tilley, N. (1997). *Realistic evaluation*. Sage Publications Ltd.

Richardson, J. E. (2007). *Analysing newspapers: An approach from critical discourse analysis*. Palgrave

Robert, L. P., Pierce, C., Marquis, L., Kim, S., & Alahmad, R. (2020). Designing fair AI for managing employees in organizations: a review, critique, and design agenda. *Human–Computer Interaction*, *35*(5-6), 545-575.

Ryan, M., & Stahl, B. C. (2021). Artificial intelligence ethics guidelines for developers and users: clarifying their content and normative implications. *Journal of Information, Communication and Ethics in Society*, *19*(1), 61-86.

Sarker, S., Chatterjee, S., Xiao, X., & Elbanna, A. (2019). The socio-technical axis of cohesion for the IS discipline: Its historical legacy and its continued relevance. *MIS Quarterly*, *43*(3), 695-719.

Schatzki, T. R. (2001). Introduction practice theory. In T. R. Schatzki, K. Knorr-Cetina, & E. Von Savigny (Eds.), *The Practice turn in Contemporary Theory* (Vol. 44, pp. 10-23). Routledge

Schiff, D., Rakova, B., Ayesh, A., Fanti, A., & Lennon, M. (2020). Principles to practices for responsible AI: Closing the gap. *arXiv*, *2006.04707*.

Schwabe, G., Katsiuba, D., Specker, R., & Dolata, M. (2025). AI-ThinkLets for brainstorming. *58th Hawaii International Conference on System Sciences (HICSS)*.

Seppälä, A., Birkstedt, T., & Mäntymäki, M. (2021). From ethical AI principles to governed AI. Proceedings of the 42nd International Conference on Information Systems (ICIS2021),

Smith, G., & Rustagi, I. (2020). Mitigating bias in Artificial Intelligence: An equity fluent leadership playbook. In: Berkeley Haas Center for Equity, Gender and Leadership.

Starke, C., Baleis, J., Keller, B., & Marcinkowski, F. (2022). Fairness perceptions of algorithmic decision-making: A systematic review of the empirical literature. *Big Data & Society*, *9*(2), 20539517221115189.

Taeihagh, A. (2021). Governance of artificial intelligence. *Policy and society*, *40*(2), 137-157.

Teodorescu, M. H., Morse, L., Awwad, Y., & Kane, G. C. (2021). Failures of fairness in automation require a deeper understanding of human-ML augmentation. *MIS Quarterly*, *45*(3).

Thornton, P. H., & Ocasio, W. (1999). Institutional logics and the historical contingency of power in organizations: Executive succession in the higher education publishing industry, 1958–1990. *American journal of Sociology*, *105*(3), 801-843.

Trist, E. L., & Bamforth, K. W. (1951). Some social and psychological consequences of the longwall method of coal-getting: An examination of the psychological situation and defences of a work group in relation to the social structure and technological content of the work system. *Human Relations*, *4*(1), 3-38.

van Dijk, T. A. (2009). Multidisciplinary CDA: A plea for diversity. In R. Wodak & M. Meyer (Eds.), *Methods of critical discourse analysis* (2nd ed., pp. 95-120). SAGE Publications

Venkatraman, N. (1989). The concept of fit in strategy research: Toward verbal and statistical correspondence. *Academy of Management Review*, *14*(3), 423-444.

Weill, P. (2008). Don't just lead, govern: How top-performing firms govern IT. *MIS Quarterly Executive*, *3*(1), 3.

Whittlestone, J., Nyrup, R., Alexandrova, A., & Cave, S. (2019). *The role and limits of principles in AI ethics: Towards a focus on tensions* ACM Conference on AI Ethics and Society,

Wirtz, B. W., Weyerer, J. C., & Sturm, B. J. (2020). The dark sides of Artificial Intelligence: An integrated AI governance framework for public administration. *International Journal of Public Administration*, *43*(9), 818-829.

Wu, P.-J. S., Straub, D. W., & Liang, T.-P. (2015). How information technology governance mechanisms and strategic alignment influence organizational performance: Insights from a matched survey of business and IT managers. *MIS Quarterly*, *39*(2), 497-518.